\documentclass{article}

\usepackage[utf8x]{inputenc}
\usepackage{float}
\usepackage[english]{babel}
\usepackage{lmodern,textcomp}
\usepackage{bm}
\usepackage{graphicx}
\usepackage{adjustbox}
\usepackage{amsmath}
\usepackage{amsmath,amsthm}
\usepackage{enumitem}
\usepackage{stmaryrd}
\SetSymbolFont{stmry}{bold}{U}{stmry}{m}{n}
\usepackage{array}
\usepackage{multirow}
\usepackage{titlesec}
\usepackage[table,xcdraw]{xcolor}
\usepackage{tikz}
\usetikzlibrary{patterns}
\usetikzlibrary{arrows.meta}
\tikzset{%
  whitebox/.style={rectangle, draw},
  lightgreybox/.style={rectangle, draw, fill=green!50},
  greybox/.style={rectangle, draw, fill=blue!30},
  redbox/.style={rectangle, draw, fill=red!40}
}
\usepackage{xspace}
\usepackage{minibox}
\usepackage{etoolbox}
\usepackage{colortbl}
\usepackage{diagbox}
\usepackage{tabularx}

\usepackage{enumitem}
\usepackage{arydshln}
\usepackage{relsize}
\usepackage[hidelinks]{hyperref}
\usepackage{listings}
\usepackage{capt-of}
\usepackage{subfiles}
\usepackage{pgfplots}
\pgfplotsset{compat=1.15}
\usepackage{ltablex}
\usepackage{pict2e}
\usepackage{pifont,dsfont}
\usepackage{amssymb}
\usepackage{hhline}
\usepackage{lmodern}

\makeatletter
\newcommand\xleftrightarrow[2][]{\ext@arrow 9999{\longleftrightarrowfill@}{#1}{#2}}
\newcommand\longleftrightarrowfill@{\arrowfill@\leftarrow\relbar\rightarrow}
\makeatother

\newcommand{\sampled}{\overset{\hspace{0.7mm} \$}{\gets}}

\newcommand{\bbracket}[1]{\llbracket #1 \rrbracket}

\newcommand{\sendright}[2]{\overset{#1}{\xrightarrow{\hspace{#2}}}}
\newcommand{\sendleft}[2]{\overset{#1}{\xleftarrow{\hspace{#2}}}}

\newcommand{\cA}{\mathcal A}
\newcommand{\cC}{\mathcal C}

\newcommand{\cK}{\mathcal K}

\newcommand{\cO}{\mathcal{O}}
\newcommand{\bbZ}{\mathbb Z}

\newcommand{\fC}{\mathfrak C}
\newcommand{\fI}{\mathfrak I}

\newcommand{\fO}{\mathfrak O}
\newcommand{\fT}{\mathfrak T}

\newcommand{\sC}{\textsf C}
\newcommand{\sE}{\textsf E}
\newcommand{\sG}{\textsf G}
\newcommand{\sR}{\textsf R}
\newcommand{\sS}{\textsf S}

\newcommand{\XOR}{\textsf{XOR}\xspace}
\newcommand{\AND}{\textsf{AND}\xspace}
\newcommand{\NOT}{\textsf{NOT}\xspace}

\newcommand{\Exp}{\textsf{Exp}\xspace}

\newcommand{\privacyC}{\mathsf{client\textsf{-}privacy}\xspace}
\newcommand{\privacyS}{\mathsf{server\textsf{-}privacy}\xspace}

\newcommand{\GC}{\textsf{GC}\xspace}
\newcommand{\Generate}{\textsf{Generate}\xspace}
\newcommand{\Encode}{\textsf{Encode}\xspace}
\newcommand{\Eval}{\textsf{Eval}\xspace}
\newcommand{\Decode}{\textsf{Decode}\xspace}

\newcommand{\OT}{\textsf{OT}\xspace}
\newcommand{\Compute}{\textsf{Compute}\xspace}

\newcommand{\AHE}{\textsf{AHE}\xspace}

\newcommand{\PKE}{\textsf{PKE}\xspace}

\newcommand{\keygen}{\textsf{KeyGen}\xspace}
\newcommand{\enc}{\textsf{Enc}\xspace}
\newcommand{\dec}{\textsf{Dec}\xspace}
\newcommand{\add}{\textsf{Add}\xspace}
\newcommand{\mulscal}{\textsf{MultScal}\xspace}
\newcommand{\mulrand}{\textsf{MultRand}\xspace}
\newcommand{\randomize}{\textsf{Randomize}\xspace}

\newcommand{\INDCPA}{\textsf{IND-CPA}\xspace}

\newcommand{\zkgen}{\textsf{ZKGen}\xspace}
\newcommand{\zkverif}{\textsf{ZKVerify}\xspace}
\newcommand{\zksim}{\textsf{ZKSim}\xspace}

\newcommand{\sk}{\textsf{sk}\xspace}
\newcommand{\pk}{\textsf{pk}\xspace}

\renewcommand{\boxtimes}{\scalebox{1.2}{ $\boxdot$ }\xspace}

\newcommand{\Keygen}{\textsf{KeyGen}\xspace}
\newcommand{\EncodeModel}{\textsf{EncodeModel}\xspace}
\newcommand{\EvalPaths}{\textsf{EvaluatePaths}\xspace}
\newcommand{\RandomizePaths}{\textsf{RandomizeScores}\xspace}
\newcommand{\EvalModel}{\textsf{EvaluateModel}\xspace}

\newcommand{\Initialization}{\textsf{InitializeGC}\xspace}
\newcommand{\ComputeScore}{\textsf{ComputeScore}\xspace}
\newcommand{\EncodeOT}{\textsf{EncodeOT}\xspace}
\newcommand{\ComputeOT}{\textsf{ComputeOT}\xspace}
\newcommand{\DecryptScore}{\textsf{DecryptScore}\xspace}
\newcommand{\ComputeResult}{\textsf{ComputeResult}\xspace}

\newcommand{\find}{\textsf{Find}\xspace}
\newcommand{\guess}{\textsf{Guess}\xspace}
\newcommand{\transcript}{\textsf{transcript}\xspace}
\newcommand{\Execute}{\textsf{Execute}\xspace}

\newcommand{\ZKP}{\textsf{ZKP}\xspace}
\newcommand{\NIZK}{\textsf{NIZK}\xspace}

\newcommand{\return}{\textsf{return}\xspace}
\newcommand{\result}{\textsf{result}\xspace}
\newcommand{\accept}{\textsf{accept}\xspace}
\newcommand{\reject}{\textsf{reject}\xspace}

\newcommand{\PreserveBackslash}[1]{\let\temp=\\#1\let\\=\temp}
\newcolumntype{C}[1]{>{\PreserveBackslash\centering}p{#1}}
\newcolumntype{R}[1]{>{\PreserveBackslash\raggedleft}p{#1}}
\newcolumntype{L}[1]{>{\PreserveBackslash\raggedright}p{#1}}

\title{Secure Decision Forest Evaluation}

\usepackage{authblk}
\author[1]{Slim Bettaieb}
\author[1,2]{Loic Bidoux}
\author[3]{Olivier Blazy}
\author[1,4,5]{Baptiste Cottier}
\author[4,5]{David Pointcheval}

\affil[1]{\small Worldline, France}
\affil[2]{\small Cryptography Research Centre, United Arab Emirates}
\affil[3]{\small XLIM, University of Limoges, France}
\affil[4]{\small DIENS, Ecole Normale Superieure de Paris, France}
\affil[5]{\small INRIA, France}
\begin{document}

\maketitle

\begin{abstract}
    Decision forests are classical models to efficiently make decision on complex inputs with multiple features. 
    While the global structure of the trees or forests is public, sensitive information have to be protected during the evaluation of some client inputs with respect to some server model.
    Indeed, the comparison thresholds on the server side may have economical value while the client inputs might be critical personal data. 
    In addition, soundness is also important for the receiver.
    In our case, we will consider the server to be interested in the outcome of the model evaluation so that the client should not be able to bias it.
    In this paper, we propose a new offline/online protocol between a client and a server with a constant number of rounds in the online phase, with both privacy and soundness against malicious clients.
\end{abstract}

\section{Introduction}

Over the past years, companies have tremendously increased the amount of data they collect from their users.
These data are often feed to machine learning algorithms in order to turn them into valuable business insights that are used to develop innovative services.
Applications include user authentication, fraud detection in banking systems, recommendation services as well as spam detection.
However, collecting and processing these data raises privacy concerns since they generally contain sensitive information regarding users.
Besides, the models used to evaluate these data may contain critical business information that also need to be protected.
In this work, we focus on the secure evaluation of decision forests which are a commonly used class of machine learning algorithms.
We consider the case where a client who holds sensitive data interacts with a server who holds a decision forest model in order to jointly evaluate the client inputs with respect to the server model.
Our goal is to ensure that the privacy of both the client and the server is guaranteed on their respective inputs.
We also investigate the scenario where the client is malicious and intends to bias the outcome of the protocol.

In order to motivate the design rationale of the proposed protocols, we consider continuous user authentication based on decision forests as an application. 
In continuous authentication, users are authenticated using a set of features that strengthen usual authentication credentials such as passwords or security tokens.
When the user's identity needs to be validated after some time interval or after some inactivity, continuous authentication offers a user-friendly experience as it avoids interrupting legitimate users and reduces the number of times they have to authenticate explicitly.
The usual scenario of continuous authentication consists of a server authenticating users based on behavioural biometrics.
A variety of features such as keystroke patterns, swiping gestures and scrolling duration are collected on the user's device and sent to the server.
The server then evaluates the client inputs with respect to its model in order to make an authentication decision.
The model is usually generated during a training step using a dedicated training dataset.

\subsection{Related Work}

Several approaches have been proposed in order to securely evaluate decision trees \cite{CCS:BPSW07, ESORICS:BFKLSS09, NDSS:BPTG15, PoPETS:WFNL16, ESORICS:TMZC17, EPRINT:DDHKNN16, PoPETS:TueKerKat19}.
However, most of these constructions are either only secure in the honest-but-curious setting or tailored for decision tree evaluation rather than decision forest evaluation.
As such, with the exception of \cite{PoPETS:WFNL16} and \cite{ESORICS:TMZC17} which we consider hereafter, the aforementioned works can't be compared to our protocol meaningfully.
Similarly to our protocol, both \cite{PoPETS:WFNL16} and \cite{ESORICS:TMZC17} rely on Additive Homomorphic Encryption ($\AHE$) and Oblivious Transfers~($\OT$).
In addition, our protocol also uses Garbled Circuits ($\GC$) in the malicious setting.

A major difference between our protocol and existing constructions is that we rely on the server sending its encrypted model to the client rather than the client sending its encrypted data to the server.
This strongly impacts the design of our protocol and allows us to introduce a preprocessing phase that can be performed offline and leveraged later during the online phrase.
This is advantageous for several use-cases such as the continuous authentication one as it offers a trade-off between the number of rounds required to execute the protocol and the required bandwidth.

All existing constructions leak some information with respect to the model structure (\emph{i.e.} on the server side) whether it be its total number of nodes $M$, the total number of comparison nodes $m$ or the maximal depth $\delta$ of the trees.
In our protocol, the client may also learn which features will be used to evaluate the trees. We consider this as a privacy leak, but as our protocols will have a complexity independent of the depth of the trees (or more precisely, the length of the paths down to the leaves), we will be able to add dummy comparisons with dummy features, which will completely hide which are the actually used features.
We thus propose two variants of our protocols in  Table \ref{F:PW}, without or with dummy comparisons.
They only differ during the offline step as the number of features in the comparisons impacts the communication and the storage: $\delta$ is the number of real features per path, whereas $\chi$ is the number of real and dummy features.
We stress that this modification does not impact the online step of our protocols: communication only depends on the number $P$ of paths, and not their length. This thus allows us to use as many dummy features as we want to hide the trees and the forest structure. This will be a crucial property for the privacy of the model.
In addition, our constructions also feature some leakage on the client side.
When used to evaluate decision forests, our protocol leaks the number of successful paths within the forest which constitutes a tolerated leakage in our targeted application.
Indeed, it corresponds to the number of accepting trees which allows to compute a confidence level associated to the result.


One can note half rounds in our constructions, which mean one-way flows, from the sender to the recipient. Indeed, most of our schemes are actually non-interactive. Since our goal is to provide the result to the server, in the malicious setting, we get a 5-flow protocol.

\begin{table}\small
\caption{Comparison to state-of-the-art \label{F:PW}}
\vspace{-\baselineskip}
\begin{center}
\begin{tabular}{|c|c|c|c|} 
  \hline
  Scheme                  & Rounds       & Tools         & Bandwidth \\ \hline \hline 
  \rowcolor{gray!30}\multicolumn{4}{|c|}{Honest-but-Curious model}  \\ \hline 
  \cite{PoPETS:WFNL16}    & 6            & $\AHE$+$\OT$  & $\cO(m)$  \\ \hline
  \cite{ESORICS:TMZC17}   & 4            & $\AHE$+$\OT$  & $\cO(m)$  \\ \hline
  \multirow{2}{*}{Section~\ref{S:HbC} (Offline)} & \multirow{2}{*}{0.5} & \multirow{3}{*}{\AHE} &$\cO(m \cdot \delta \cdot  2^\nu)~$ \\ \cline{4-4}
  & &  & $\cO(m \cdot \chi \cdot  2^\nu)$ \\ \cline{1-2} \cline{4-4}
  Section~\ref{S:HbC} (Online) & 0.5 & &$\cO(P)$ \\ \hline \hline

  \rowcolor{gray!30}\multicolumn{4}{|c|}{Malicious model} \\ \hline 
  \cite{PoPETS:WFNL16}  & 2 & $\AHE$+$\OT$ & $\cO(M)$ \\ \hline
  \cite{ESORICS:TMZC17} & 4 & $\AHE$+$\OT$ & $\cO(m)$ \\ \hline
  \multirow{2}{*}{Section~\ref{S:Malicious} (Offline)} & \multirow{2}{*}{0.5} & \multirow{2}{*}{\AHE} & $\cO(m \cdot \delta \cdot  2^\nu)$ \\ \cline{4-4}
  & & & $\cO(m \cdot \chi \cdot  2^\nu)$ \\ \cline{1-4} 
  Section~\ref{S:Malicious} (Online) & 2.5 & $\AHE$+$\GC$+$\OT$ & $\cO(P)$ \\ \hline 
\end{tabular} \vspace*{-2em}
\end{center}
\end{table}

\subsection{Contributions}

In this paper, we propose several constructions to securely evaluate decision forests with binary output classes.
In our setting, a server evaluates a model $\mathcal{M}$ with respect to some client inputs $\bm x$ and \accept or \reject~the client according to the evaluation outcome.
We consider the server to be honest-but-curious and describe two protocols that are respectively tailored for honest-but-curious and malicious clients.
As we target applications where the interactions between the client and the server should be as low as possible, we design two-step protocols in which some part of the computation can be performed offline, and the communication performed before knowing the inputs of the client.

In the honest-but-curious setting, our protocol only requires one flow from the client to the server to be executed during the online step which outperforms previous results from the literature.
In the malicious setting, our protocol can be seen as a trade-off between existing constructions with respect to the number of rounds required and the bandwidth cost.
However, our protocols leak the number of trees successfully evaluated within the forest to the server.
We consider this as a feature rather than a drawback as we want the server to learn the outcome of the evaluation in our setting.
Indeed, this additional information is generally used as a confidence score with respect to the evaluation outcome and is expected to be known in some use-cases such as the continuous authentication one.

Another contribution is the fact that in our context, we use garbled circuit in a malicious setting, without any additional techniques compared to the honest-but-curious setting. Thereby, we do not need to use solutions such as \emph{Cut \& Choose}. Even if this solution has been well studied and optimized \cite{PKC:MohFra06b,EC:LinPin07,C:Lindell13,C:LinRiv14,EC:AMPR14,EC:WanMalKat17}, this is still an expensive solution requiring $\ell$ garbled circuits for statistical security $2^{-\ell}$. In a survey, Dupin et al.~\cite{PROVSEC:DupPoiBid18} showed that a malicious generator can corrupt a garbled circuit only by adding \NOT-gates or by failure attacks, but we can protect against these  attacks in our context. Indeed, in case of failure attacks, the adversary will likely be rejected, without learning any information about the thresholds. In the case of wrong circuit (with additional \NOT-gates), we have introduced random inversions of the outcomes of the paths, and so an attack will reduce to guess all (or most of) the inversions, which will again likely lead to a reject, without leaking any information.

\subsection{Paper Organization}

We present the main tools that will be used to design our secure decision forest evaluation protocols in Section \ref{S:Preliminaries}, 
Then, we describe and analyze our construction in the honest-but-curious setting in Section~\ref{S:HbC}.
Eventually, we move to the malicious setting in Section~\ref{S:Malicious}.
Performances and results for several applications are discussed in Section~\ref{S:Efficiency}.

\section{Preliminaries}
\label{S:Preliminaries}

\subsection{Decision Tree Learning}

Decision tree learning is a discipline used to solve multi-criteria problems that can be modelled using \emph{decision trees}.
In this paper, we focus on binary classification trees, which are decision trees whose leaves can take two values (the two output classes).
In order to improve the accuracy of the model, one often considers \emph{decision forests} (sets of decision trees) where each tree of the forest is evaluated separately and then aggregated for the final decision.

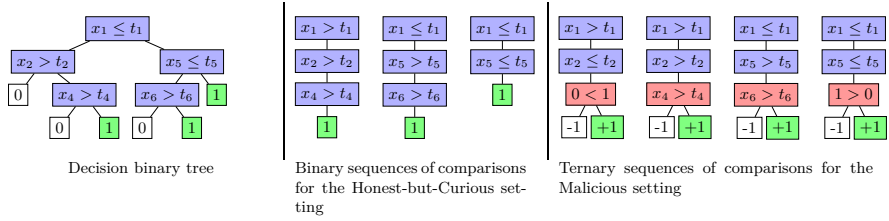
\begin{figure*}\small
	\begin{center}
		\resizebox{\textwidth}{!}{
		\begin{tabular}{c|ccc|cccc}
			\parbox[t][][b]{16em}{
				\begin{tikzpicture}
					[level distance = 2em,
					level 1/.style = {sibling distance = 3em}]
					\node [greybox] {$x_1 \leq t_1$}
						child {node [greybox, xshift=-3em] {$x_2 > t_2$}
							child {node [whitebox] {0}}
							child {node [greybox, xshift=1em] {$x_4 > t_4$}
								child {node [whitebox] {0}}
								child {node [lightgreybox] {1}}
							}
						}
						child {
							node [greybox, xshift=3em] {$x_5 \leq t_5$}
							child {node [greybox] {$x_6 > t_6$}
								child {node [whitebox] {0}}
								child {node [lightgreybox] {1}}}
							child {node [lightgreybox] {1}}
							};
				\end{tikzpicture}}
		&
			\parbox[t][][b]{4em}{
				\begin{tikzpicture}
					[level distance = 2em]
					\node [greybox] {$x_1 > t_1$}
						child {node [greybox] {$x_2 > t_2$}
							child {node [greybox] {$x_4 > t_4$}
                                child {node [lightgreybox] {1}}}};
				\end{tikzpicture}}
		&
			\parbox[t][][b]{4em}{
				\begin{tikzpicture}
					[level distance = 2em]
					\node [greybox] {$x_1 \leq t_1$}
						child {node [greybox] {$x_5 > t_5$}
							child {node [greybox] {$x_6 > t_6$}
                                child {node [lightgreybox] {1}}}};
				\end{tikzpicture}}
		&
			\parbox[t][][b]{4em}{
				\begin{tikzpicture}
					[level distance = 2em]
					\node [greybox] {$x_1 \leq t_1$}
						child {node [greybox] {$x_5 \leq t_5$}
                            child {node [lightgreybox] {1}}};
				\end{tikzpicture}}
		&
			\parbox[t][][b]{4em}{
				\begin{tikzpicture}
					[level distance = 2em,
					level 1/.style = {sibling distance = 2em}]
					\node [greybox] {$x_1 > t_1$}
						child {node [greybox] {$x_2 \leq t_2$}
							child {node [redbox] {$0 < 1$}
								child {node [whitebox] {-1}}
								child {node [lightgreybox] {+1}}}};
				\end{tikzpicture}}
		&
			\parbox[t][][b]{4em}{
				\begin{tikzpicture}
				[level distance = 2em,
				level 1/.style = {sibling distance = 2em}]
				\node [greybox] {$x_1 > t_1$}
					child {node [greybox] {$x_2 > t_2$}
						child {node [redbox] {$x_4 > t_4$}
							child {node [whitebox] {-1}}
							child {node [lightgreybox] {+1}}}};
			\end{tikzpicture}}
		&
			\parbox[t][][b]{4em}{
				\begin{tikzpicture}
				[level distance = 2em,
				level 1/.style = {sibling distance = 2em}]
				\node [greybox] {$x_1 \leq t_1$}
					child {node [greybox] {$x_5 > t_5$}
						child {node [redbox] {$x_6 > t_6$}
							child {node [whitebox] {-1}}
							child {node [lightgreybox] {+1}}}};
			\end{tikzpicture}}
		&
			\parbox[t][][b]{4em}{
				\begin{tikzpicture}
				[level distance = 2em,
				level 1/.style = {sibling distance = 2em}]
				\node [greybox] {$x_1 \leq t_1$}
					child {node [greybox] {$x_5 \leq t_5$}
						child {node [redbox] {$1 > 0$}
							child {node [whitebox] {-1}}
							child {node [lightgreybox] {+1}}}};
			\end{tikzpicture}} \\ &&&&&&& \\
			\multicolumn{1}{c}{Decision binary tree}
			& \multicolumn{3}{p{14em}}{Binary sequences of comparisons for the Honest-but-Curious setting}
			& \multicolumn{4}{p{19em}}{Ternary sequences of comparisons for the Malicious setting}\\[-1em]
		\end{tabular}
		}
	\caption{Decision tree, Sequences of comparisons, Binary sequences \label{Fig:Tree}}
\end{center}\vspace*{-1em}
\end{figure*}

A decision forest is thus a list of binary decision trees, as one is shown on Figure~\ref{Fig:Tree}, on the left part. They can each be converted into a list of comparison sequences down to accepting leaves (in the center of Figure~\ref{Fig:Tree}). Each comparison is between a feature value $x_i$ and the threshold value $t_i$ from the model: we thus denote the model $\mathcal{M} = (P, \delta, \nu, \tau, (t_{i, j}, v_{i, j})_{i \in [P], j \in [\delta]})$ to represent a binary decision forest and $\bm x = (x_{i, j})_{i \in [P], j \in [\delta]}$ the inputs of size $\nu$ to be evaluated.
The model $\mathcal{M}$ represents $P$ paths of maximal depth $\delta$ that can be evaluated to compute a score in order to determine the output of the evaluation with respect to some threshold $\tau$ on the number of accepted paths (which is equal to the number of accepting trees).
Each path is a series of comparisons, where the $(i,j)$-th comparison denotes the comparison of depth $j$ in the $i$-th path: namely, the comparison of input $x_{i,j}$ and threshold $t_{i,j}$.
In addition, the boolean values $v_{i, j}$ are used to determine the comparison operator.  \emph{Lower or equal} $(\leq)$ whenever $v_{i,j} = 1$ or \emph{Strictly greater} $(>)$ whenever $v_{i,j} = 0$.
A path is considered to be accepting if all its comparisons yields to \texttt{TRUE}, otherwise it is rejecting.
And then a tree is accepting if one path is accepting. One can note that in the sequences extracted from a tree, at most one is accepting.
Given some input $\bm x = (x_{i, j})_{i \in [P], j \in [\delta]}$, the number of accepting paths with respect to model $\mathcal{M}$ is denoted by $\mathcal{M}(\bm x)$. It then corresponds to the number of accepting trees.
Hence, the outcome of the decision forest evaluation is the Boolean $(\mathcal{M}(\bm x) \geq \tau)$.

More concretely, as shown on Figure~\ref{Fig:Tree}, if we have a forest with $T$ trees, we extract all the paths down to accepting leaves, with the successive comparisons $(t_{j}, v_{j})_j$. On a given input $\bm x$, each tree has at most one accepting path, and so at most $T$ accepting paths in total. Then, we can decide to accept $\bm x$ when at least $T/2$ among the $P$ paths (the majority of the trees) are accepting: $\tau = T/2$.
We stress that in this scenario, each path will be considered accepting or rejecting. But we will just expect at least $T/2$ accepting paths among $P$. This will be enough for our protocol in the honest-but-curious setting: we essentially ignore rejecting paths.

For the malicious setting, we will consider more complex sequences of comparisons, with a ternary output: accepting, rejecting, or ignoring. This is the right part of Figure~\ref{Fig:Tree}: if the last red comparison is reached, a decision is taken, as accept (+1) or reject (-1), whereas when the last comparison is not reached, the path will be ignored (0). This way, most of the paths will be ignored, and exactly one path will be accepting or rejecting, as the global tree would be. We stress that some 'always true' or 'always false' comparisons will have to be added to make the above technique work properly.
This is easy to see that the three ways of representing and evaluating a binary decision tree are equivalent, with a final outcome 'accept' or 'reject'. The last one will allow to prevent malicious behaviours from the client, in order to falsely get accepted.


\subsection{Public-Key Encryption}
A \emph{public-key encryption} scheme $\PKE$ is defined by three algorithms $(\keygen, \enc, \dec)$:
\begin{itemize}
  \item $\keygen(1^\kappa)$: with input $\kappa$ as security parameter, returns a public encryption key $\pk$ and a private decryption key $\sk$.
  \item $\enc(\pk, m)$: returns $\bbracket{m}$, an encryption of $m$ under the public encryption key $\pk$ 
  \item $\dec(\sk, \bbracket{m})$: returns $m$.
\end{itemize}
Such an encryption scheme should provide secrecy of the message. But as the encryption key is public, anybody can encrypt any message of its choice. We thus talk about \emph{indistinguishability against chosen-plaintext attacks} (\INDCPA).


\subsection{Homomorphic Encryption}
An \emph{Additively Homomorphic Encryption scheme} \AHE on plaintexts over an additive group of size $p$ (typically, it will be $\bbZ_p$) is a \PKE scheme with two more algorithms ($\add$, $\mulscal$):
\begin{itemize}
  \item $\add(\pk, \bbracket{m_1},\bbracket{m_2})$: Given $\pk$ and two ciphertexts $\bbracket{m_1}, \bbracket{m_2}$, returns  $\bbracket{m_1} \boxplus \bbracket{m_2} = \bbracket{m_1 + m_2}$ an encryption of the sum of the plaintexts under the same public key $\pk$;
  \item $\mulscal(\pk, \bbracket{m}, k)$: Given $\pk$, a ciphertext $\bbracket{m}$, and a scalar $k\in\bbZ_p$, returns $k \boxtimes \bbracket{m} = \bbracket{k \cdot m}$;
\end{itemize}
Two randomization properties will also be considered:
\begin{itemize}
  \item $\randomize(\pk,\bbracket{m})$: Given $\pk$ and a ciphertext $\bbracket{m}$, returns a different ciphertext of $m$. It can be implemented as $\add(\pk, \bbracket{m}, \enc(\pk, 0))$;
  \item $\mulrand(\pk,\bbracket{m})$: Given $\pk$, a ciphertext $\bbracket{m}$, returns a ciphertext of $k\cdot m$, for a non-zero random $k$.
\end{itemize}

\paragraph{Lifted ElGamal Encryption Scheme.} ElGamal~\cite{C:ElGamal84} encryption is a multiplicative homomorphic encryption scheme. To make it additive, we encode a message as $g^m$. Also, after decryption, we retrieve $g^m$. No discrete logarithm computation is required as we only check if $m$ belongs to a given interval $[\tau_{\min}, \tau_{\max}]$, which can be done by checking either $g^m \in \{g^i\}_{i \in [\tau_{\min}, \tau_{\max}]}$ or not. 
\begin{itemize}
  \item $\keygen(1^\kappa)$: Generates a cyclic group $G$ of order $p$ with $|p|=\kappa$, with generator $g$; samples $x \sampled \bbZ_p$; returns $\pk=(G,p,g,h=g^x)$ and $\sk=x$;
  \item $\enc(\pk, m)$: Generates $y \sampled \bbZ_p$; computes $c_1=g^y$ and $c_2=g^m \cdot h^y$; returns $\bbracket{m}=(c_1,c_2)$;
  \item $\dec(\sk, \bbracket{m}=(c_1,c_2))$: Computes $c_2 \cdot c_1^{-x}=g^m$; 
  \item $\add(\pk, \bbracket{m_1}=(c_1^1,c_2^1), \bbracket{m_2}=(c_1^2,c_2^2))$: Computes $c_1^3=c_1^1 \cdot c_1^2$ and $c_2^3=c_2^1 \cdot c_2^2$; returns $\bbracket{m_1+m_2}=(c_1^3,c_2^3)$;
  \item $\mulscal(\pk, \bbracket{m}=(c_1, c_2), k)$: Computes $c'_1=(c_1)^k$ and $c'_2=(c_2)^k$; returns $\bbracket{k \cdot m}=(c'_1,c'_2)$;
  \item \randomize and \mulrand are computed thanks to \add and \mulscal as described above.
\end{itemize}

\subsection{Oblivious Transfer}
An \emph{Oblivious Transfer} $\OT$ is a two-party protocol between a sender with input a pair of messages $(m_0, m_1)$ and a receiver with input a bit $b$ that allows the receiver to retrieve $m_b$.
The receiver should not learn anything about $b$, while the receiver should not learn anything about $m_{1-b}$. 
A particular family of $\OT$ can be defined as a tuple of algorithms $(\Encode, \Compute, \Decode)$:
\begin{itemize}
  \item $\Encode(b)$: Given a bit $b$, returns the encoded value $\widetilde{b}$;
  \item $\Compute((m_0,m_1),\widetilde b)$: Given two messages $(m_0, m_1)$ and an encoded value $\widetilde{b}$, returns the encoding $\widetilde m$ associated to $m_b$;
  \item $\Decode(\widetilde m)$: Given $\widetilde m$, returns the message $m_b$.
\end{itemize}

\paragraph{Security Properties.}
Two main security notions are expected: the sender-privacy, which hides $m_{1-b}$ to the receiver, and the receiver-privacy, which hides $b$ to the sender. 
We will focus on two different cases: receiver-privacy against a malicious sender; and sender-privacy against an honest-but-curious receiver.
This will be enough for our application to decision trees, where the server (receiver) will be considered honest-but-curious while the client will possibly behave maliciously (sender).

In particular, Even-Goldreich-Lempel~\cite{C:EveGolLem82} proposed such an efficient oblivious transfer from any \INDCPA public-key encryption scheme \PKE, with an efficient uniform sampling algorithm in the set of the public keys $\cK$. It is secure against a malicious sender and an honest-but-curious receiver.

\subsection{Garbled Circuits}
A \emph{Garbled Circuit} \cite{FOCS:Yao86} is a primitive that allows two parties, a generator and an evaluator, to jointly compute a function over their respective private inputs.
The computation to be performed must be modelled by a Boolean \emph{circuit} using logic gates.
Hereafter, we only consider \AND or \XOR logic gates namely gates with two input bits and one output bit.
In the basic form of Garbled Circuits, for each logic gate, the generator generates a pair of random symmetric keys $(k_0, k_1)$ for each input or output bit where $k_0$ and $k_1$ are respectively associated to the bit values $0$ and $1$.
Each gate can be encoded (or \emph{garbled}) using a symmetric encryption scheme by generating four ciphertexts where each ciphertext encrypts the output key corresponding to one output of the logic gate under the corresponding input keys.
In practice, one can use the point-and-permute \cite{STOC:BeaMicRog90}, free-\XOR \cite{ICALP:KolSch08} and half-gate~\cite{EC:ZahRosEva15} optimizations in order to reduce the number of ciphertexts, and even avoid the use of symmetric encryption with only hash functions.
Given all the input keys corresponding to its input bits, the evaluator can recursively get the output key of the last gate thus retrieving the Boolean circuit outcome.



While the generator knows the input keys corresponding to its input $\alpha$, and can then provide them to the evaluator with all the ciphertexts, it does not know the input $\beta$ of the evaluator. The input keys corresponding to that input $\beta$ are obtained using Oblivious Transfer, between the generator as the sender, and the evaluator as the receiver.
More details can be found in the appendix.

\subsection{Secure Equality Test}
Our protocol relies on a secure equality test in the malicious client setting.
A garbled circuit testing the equality between two $\kappa$-bit values $\alpha$ and $\beta$ can be computed by
\begin{align*}
	(\alpha = \beta) 	&= \left(\alpha[\ell] = \beta[\ell], \forall \ell \in  [\kappa]\right) = \left(\alpha[\ell] \oplus \beta[\ell] = 0, \forall \ell \in  [\kappa]\right) \\
					  &= \left(\bigwedge_{\ell=1}^\kappa (\overline{\alpha[\ell] \oplus \beta[\ell]})=1 \right) \\
						&= \left(\bigwedge_{\ell=1}^\kappa (\overline{\alpha[\ell]} \oplus \beta[\ell])=1\right)
\end{align*}

Each equality test requires $\kappa$ \XOR-gates of arity 2 and a global \AND-gate of arity $\kappa$, or $\kappa-1$ \AND-gates of arity 2.
Using the free-\XOR and half-gate optimizations, such an equality test can be computed using $2(\kappa - 1)$ ciphertexts. 
One may compare hashed values of $\alpha$ and $\beta$, of shorter length, at the cost of possibly false positive cases.



\subsection{Zero-Knowledge Proofs} 
A \emph{Zero-Knowledge Proof} (\ZKP) is a protocol between a prover $P$, who wants to prove to a verifier $V$, that a given statement belongs to a language, without leaking any information about the witness. It can be made non-interactive (\NIZK). Such a proof must be sound, which means that no adversary can generate an acceptable proof when $x \notin L$, but with negligible probability; and zero-knowledge, which guarantees zero-leakage about the witness. More details can be found in the appendix.

\section{Honest-but-Curious Client and Server}
\label{S:HbC}

In this section, we describe a protocol providing secure decision forests evaluation in the case where both the client and the server are honest-but-curious.
In this setting, participants genuinely follow the protocol but may attempt to learn information from legitimately received messages.
The protocol allows to evaluate the client inputs $\bm x = (x_{i, j})_{i \in [P], j \in [\delta]}$ with respect to the server model represented as binary sequences of comparisons $\mathcal{M} = (P, \delta, \nu, \tau, (t_{i, j}, v_{i, j})_{i \in [P], j \in [\delta]})$.
The client should not learn anything on the threshold $\tau$ or the comparisons performed by the model which are described using $t_{i, j}$ and $v_{i, j}$ respectively.
The server should not learn anything regarding the client inputs $x_{i, j}$.
Ideally, the server should only learn the outcome of the evaluation but we tolerate the leakage of the number of paths successfully evaluated by the model so that the server can make a decision according to the threshold $\tau$ on the number of successful paths. We have already noted this corresponds to the number of accepting trees, which helps to get a confidence score for the decision.

\subsection{Protocol Description} \label{S:HbCProt}
\begin{figure*}[!ht]
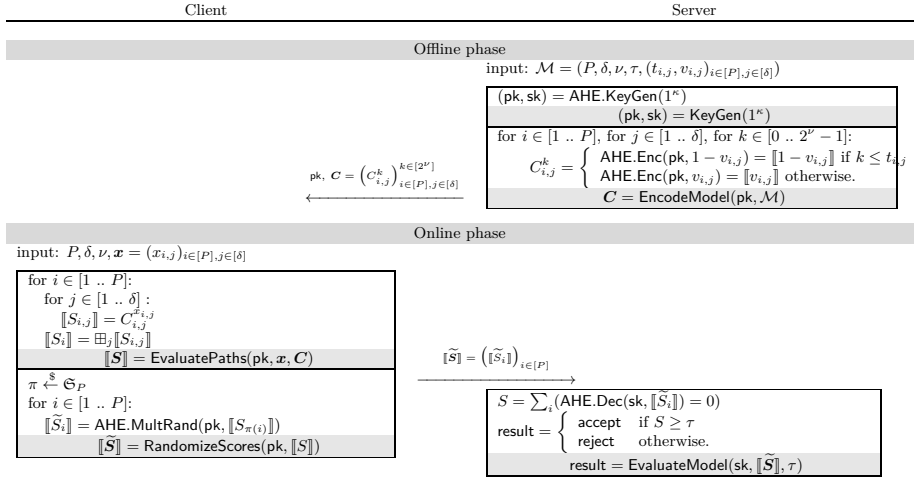
\small
	\begin{center}
		\resizebox{\textwidth}{!}{
	\begin{tabular}{lp{3em}l}
	\multicolumn{1}{c}{Client}
		&& \multicolumn{1}{c}{Server} \\ \hline \\
	\rowcolor{gray!30} \multicolumn{3}{c}{Offline phase} \\
	&& input: $\mathcal{M} = (P, \delta, \nu, \tau, (t_{i, j}, v_{i, j})_{i \in [P], j \in [\delta]})$ \\[.5em]

		&& \begin{tabular}{|p{24em}|} \hline
	    $(\pk,\sk) = \AHE.\keygen(1^\kappa)$ \\
		\multicolumn{1}{|c|}{\cellcolor{gray!20} $(\pk, \sk) = \Keygen(1^\kappa)$} \\ \hline
		\end{tabular} \\[.5em]
	
		&& \begin{tabular}{|p{24em}|} \hline
			for $i \in [1 \ .. \ P]$, for $j \in [1 \ .. \ \delta]$, for $k \in [0\ .. \ 2^\nu-1]$: \\
			\qquad $C_{i,j}^k = \left\{
				\begin{array}{l}
					\AHE.\enc(\pk,1-v_{i,j}) = \bbracket{1-v_{i,j}} \mbox{ if } k \leq t_{i,j} \\
					\AHE.\enc(\pk,v_{i,j}) = \bbracket{v_{i,j}} \mbox{ otherwise.}
				\end{array}
				\right.$ \\
        \multicolumn{1}{|c|}{\cellcolor{gray!20} $\bm C = \EncodeModel(\pk, \mathcal{M})$} \\ \hline
			\end{tabular} \\[-3em]
    \multicolumn{2}{r}{$\sendleft{\pk,~\bm C\;=\;\big(C_{i,j}^k\big)_{i \in [P], j \in [\delta]}^{k \in [2^\nu]}}{3cm}$} \\[1em]
	\rowcolor{gray!30} \multicolumn{3}{c}{Online phase} \\
	input: $P, \delta, \nu, \bm x = (x_{i,j})_{i \in [P], j\in[\delta]}$ \\[.5em]

		\begin{tabular}{|p{22em}|} \hline
			for $i \in [1 \ .. \ P]$: \\
	    	\quad for $j \in [1 \ .. \ \delta]: $\\
			$\qquad \bbracket{S_{i,j}}=C_{i,j}^{x_{i,j}}$ \\
	    	\quad $ \bbracket{S_i}=\boxplus_j \bbracket{ S_{i,j}}$ \\
			\multicolumn{1}{|c|}{\cellcolor{gray!20} $\bm{\bbracket S} = \EvalPaths(\pk, \bm x, \bm C)$} \\ \hline
		\end{tabular} \\

		\begin{tabular}{|p{22em}|} \hline
			$\pi \sampled \mathfrak S_{P}$ \\
	    	for $i \in [1 \ .. \ P]$: \\
      		\quad $\bbracket{\widetilde S_i}=\AHE.\mulrand(\pk,\bbracket{S_{\pi(i)}})$ \\
			\multicolumn{1}{|c|}{\cellcolor{gray!20} $\bbracket{\widetilde{\bm S}} = \RandomizePaths(\pk,\bbracket{S})$} \\ \hline
		\end{tabular} \\[-7em]
		& \multicolumn{2}{l}{$\sendright{\bbracket{\widetilde{\bm S}}\;=\;\big(\bbracket{\widetilde S_i}\big)_{i \in [P]}}{3cm}$} \\

		&& \begin{tabular}{|p{23em}|} \hline
			$S = \sum_i (\AHE.\dec(\sk,\bbracket{\widetilde S_i})=0)$ \\ 
			$\result = \left\{
				\begin{array}{ll}
					\accept & \mbox{if } S \geq \tau \\
					\reject & \mbox{otherwise.}
				\end{array}
				\right.$ \\
			\multicolumn{1}{|c|}{\cellcolor{gray!20} $\result = \EvalModel(\sk, \bbracket{\widetilde{\bm S}}, \tau)$} \\ \hline
			\end{tabular}
	\end{tabular}}
	\end{center}
    \caption{Secure decision forest evaluation for Honest-but-Curious Client and Server \label{Fig:HbC}}
  \end{figure*}

As illustrated on Figure~\ref{Fig:HbC}, our protocol can be seen as a tuple of algorithms (\Keygen, \EncodeModel, \EvalPaths, \RandomizePaths, \EvalModel) where \Keygen, \EncodeModel and \EvalModel are computed by the server while \EvalPaths and \RandomizePaths are computed by the client.
The $\EncodeModel$ algorithm is a preprocessing step that returns an encoded model $\bm C$ from the server secret key $\sk$ and the model $\mathcal{M}$.
The encoded model $\bm C$ is used along with the public key $\pk$ and client inputs $\bm x$ by the $\EvalPaths$ algorithm in order to compute the encoded scores $\bm S$ of each path of the model.
Next, these encoded scores are randomized and permuted by the $\RandomizePaths$ algorithm which outputs the randomized scores $\widetilde{\bm S}$.
The server ends the protocol by computing the $\EvalModel$ algorithm that takes the secret key $\sk$, the randomized scores $\widetilde{\bm S}$ and the threshold $\tau$ as inputs and returns the outcome of the evaluation of $\bm x$ with respect to $\mathcal{M}$.

During the $\EncodeModel$ preprocessing step, a ciphertext $C_{i,j}^k$ is computed for each comparison node $(i, j)$ of the model (where $i$ is the index of the path and $j$ the depth of the node) and each possible input value $k \in [2^\nu]$ as follows:\\
$C_{i,j}^k=\left\{\begin{array}{rll}                                                                                      \textsf{AHE.Enc}(\pk,1-v_{i,j}) & =\bbracket{1-v_{i,j}} & \mbox{if } k \leq t_{i,j} \\
	\textsf{AHE.Enc}(\pk,v_{i,j}) & = \bbracket{v_{i,j}} & \mbox{otherwise.}                                              
\end{array} \right. $ \\ As $v_{i, j} = 1$ when the expected comparison result is $(x_{i,j} \leq t_{i,j})$ and $v_{i,j} = 0$ otherwise, $C_{i,j}^k$ is a ciphertext of $0$ (respectively a ciphertext of $1$) whenever the input value $k$ satisfies (respectively does not satisfy) the comparison.
One can see that the number of $C_{i,j}^k$ ciphertexts is exponential with respect to $\nu$ but we stress that one only needs a few bits of precision in order to get a meaningful outcome. This number is thus linear in the number of comparisons in practice.
During the $\EvalPaths$ step, the client retrieves the ciphertexts $C_{i,j}^{x_{i,j}}$ using its inputs $x_{i,j}$ and uses them to compute the encrypted score $\bbracket{S_i}$ of each path $i$.
Such scores are ciphertexts of $0$ if all the comparisons of the path are successful and ciphertexts of a non-zero value otherwise.
We indeed stress that the ciphertexts encode the negation of the result of the comparison: 0 if true and 1 if false. As soon a false comparison happens, the sum $S_i$ becomes non-zero.

The $\RandomizePaths$ step guarantees client privacy by randomizing and permuting the encrypted scores $\bbracket{S_i}$ without altering the fact that $S_i = 0$ if the path $i$ is successful.
During the $\EvalModel$ step, the server decrypts the randomized scores $\bbracket{\widetilde S_i}$ in order to retrieve the number of successful paths and returns the outcome of the model evaluation with respect to the threshold $\tau$.

\subsection{Protocol Security}
\label{S:HbCProof}

\paragraph{Correctness and Soundness.}
The correctness directly follows from the construction of the ciphertexts. 
If all the comparisons are correct, each $S_{i,j}$ is equal to $0$ and the sum $S_i$ is $0$ (correctness), otherwise, at least one $S_{i,j}$ equals $1$ and $S_i$ does not equal $0$ (soundness). 
After the client randomizes the $S_i$, zeros are still zeros, while other $S_i$ become random values. When decrypting, the server counts the zeros. The threshold $\tau$ is applied for the final decision.

\paragraph{Client Privacy.}
An honest-but-curious server should not learn any client secret information, except what it can learn from the outcome: the number of successful paths (See Figure~\ref{Fig:HbC_games}, on the left).

We thus consider an adversary against the privacy of the client: it first chooses a model $\mathcal{M}$ for the server, and two sets of possible inputs $(\bm x_0,\bm x_1)$ for the client.
It also provides the random tape $\rho$ of the server. 
The adversary sees the transcript between a server using $\mathcal{M}$ and $\rho$, and a client using ${\bm x}_b$ for a random bit $b$, and it should guess $b$. 
There is the natural restriction that $\mathcal{M}(\bm x_0) = \mathcal{M}(\bm x_1)$. The random tape $\rho$ will be used by the server for encoding the model $\mathcal{M}$.
\begin{figure*}\small
\begin{center}
\fbox{\begin{minipage}{.45\textwidth}
Experiment $\Exp_\cA^{\privacyC-b}(\kappa,\cA)$:
\begin{enumerate}
  \item $((\bm x_0,\bm x_1), (\mathcal{M}, \rho)) \gets  \mathcal A.\find()$
  \item $\transcript \gets  \Execute(({\bm x}_b),(\mathcal{M}, \rho))$
  \item $b' \gets  \cA.\guess(\transcript)$
  \item if $\mathcal{M}(\bm x_0) = \mathcal{M}(\bm x_1)$ \return $(b' = b)$ \\
  \qquad otherwise \return a random bit
\end{enumerate}
\end{minipage}}
\fbox{\begin{minipage}{.45\textwidth}
Experiment $\Exp_\cA^{\privacyS-b}(\kappa)$:
\begin{enumerate}
  \item $((\bm x, \rho), (\mathcal{M}_0, \mathcal{M}_1)) \gets  \cA.\find()$
  \item $\transcript \gets \Execute((\bm x, \rho), (\mathcal{M}_b))$
  \item $b' \longleftarrow \cA.\guess(\transcript)$
  \item if $(\mathcal{M}_0(\bm x) \geq \tau_0) = (\mathcal{M}_1(\bm x) \geq \tau_1)$ \return $(b' = b)$ \\
  \qquad otherwise \return a random bit
\end{enumerate}
\end{minipage}}
\caption{Client-Privacy Security Game (left) and Server-Privacy Security Game (right), in the Honest-but-Curious Setting \label{Fig:HbC_games}}
\end{center}
\end{figure*}

For our scheme, the client privacy is provided thanks to the permutation and randomization of the encrypted scores: from the expected outcome, one can encrypt the correct number of 0, and the other values are non-zero random values. One can then randomize and permute them. This is indistinguishable from the server point of view.

\paragraph{Server Privacy.} 
An honest-but-curious client should not learn any server secret information, except what it can learn from the outcome namely the accept or reject decision (See Figure~\ref{Fig:HbC_games}, on the right).

Hence, we consider an adversary that chooses some inputs $\bm x$ for the client and its random tape $\rho$, but two different models $\mathcal{M}_0 = (P, \delta, \nu, {\bm t}_0, {\bm v}_0, \tau_0)$ and $\mathcal{M}_1 = (P, \delta, \nu, {\bm t}_1, {\bm v}_1, \tau_1)$, with the constraint that evaluating $\bm x$ with respect to the two models $\mathcal{M}_0$ and $\mathcal{M}_1$ should produce the same result.
The random tape $\rho$ will be used by the client for randomizing the ciphertexts.
The adversary should then distinguish transcripts involving the two models.

For our scheme, the server's privacy is provided by the encryption of the model in $\bm C$, during the offline phase.
The private server's information are the thresholds $t_{i,j}$ and the Boolean values $v_{i,j}$ for each comparison as well as the final threshold $\tau$.
One can note that the client learns which feature is used in a given comparison. 
This can be avoided by adding dummy comparisons so that each feature (or many features) is used in every path as discussed previously.
For the formal proof, as the scheme leaks the number of accepting paths, this is used by the simulator for the final outcome, without needing the decryption key. Then, as the decryption key is not known anymore, using \INDCPA, we can replace all the ciphertexts in the offline phase by encryptions of 0: the client cannot learn anything anymore.

\section{Malicious Client and Honest-but-Curious Server}
\label{S:Malicious}

Unfortunately, a malicious client could trivially bias the outcome of the protocol described in Figure \ref{Fig:HbC} by setting all the $\bbracket{\widetilde S_i}$ as encryptions of zeros so that he will be accepted by the server independently of its inputs. He knows accepting paths should encrypt 0, he can force that in his unique flow to the server.

In this section, we describe a protocol providing secure decision forests evaluation even if the client behaves maliciously in order to get accepted, while the server is still honest-but-curious.
Our security goals remain unchanged from Section \ref{S:HbC}, however the client may now deviate from the protocol to influence the evaluation outcome.
In order to secure our protocol, we add some randomness within the model through the notion of \emph{path polarity} and rely on secure equality tests.

\subsection{Protocol Description} \label{S:MalProt}
In order to avoid the above attack, we introduce the notion of path polarity $p_i$: the client cannot predict anymore the expected outcome of a path. With thus now use an enriched model $\mathcal{M} = (P, \delta, \nu, \tau, (t_{i,j}, v_{i,j}, p_i)_{i \in [P], j \in [\delta]})$, with ternary sequences of comparisons (as shown of Figure~\ref{Fig:Tree}) with a polarity $p_i$. Indeed, to be able to exploit the polarity, a path should have three possible outcomes: 'accept', 'reject', 'ignore'.
Then, when a path is positive ($p_i = 1$), according to the computed value, -1, 0, or +1, it will be 'reject', 'ignore', or 'accept', respectively; if the path is negative ($p_i = -1$), according to the computed value, -1, 0, or +1, it will be 'accept', 'ignore', or 'reject', respectively.
From a binary tree, such a path is now the path down to the last node that has two distinct leaves: on an input $\bm x$, if it does not reach the last node (some of the comparisons fails before), one outputs 0, otherwise one outputs -1 or +1, whether the leaf is rejecting or accepting. A tree of depth $\delta$ has at most $2^{\delta-1}$ such disjoint paths: an input $\bm x$ must be accepted/rejected by exactly one path only, all the other paths should output 'ignore'.
It is possible to extract such paths from any binary decision tree, by possibly adding some 'always true'/'always false' nodes.
Such 'always true'/'always false' comparisons will also be added to hide the actually used features. This will lead to an impossibility for the malicious client to guess the outcome of a path from the features used in the comparisons: the client does not know if the comparison is really exploited, and the client does not know the polarity and thus whether it should force +1 or -1 to increase the score.

Intuitively, the best attack of the adversary is by guessing the polarity of the paths to hope to pass the threshold. But as the polarity is random and hidden (as no information leaks, as proven later), the sum of the outputs of a malicious client will follow a binomial distribution with bias 1/2. And the expected sum is 0. If we set the threshold not too low, the probability to get accepted is negligible. An alternative is also to add some 'always accepting' paths, to artificially increase the expected sum of an honest user. With 20 such paths, we can set $\tau = 20$.
Let us thus consider 120 paths with a threshold $\tau=20$ (with either additional 'always accepting' paths, or an initially high threshold): to pass the threshold, one needs 20 correct guesses (probability bounded by $1/10^6$, on exactly 20 non-zero outputs), or at least 70 successes among 120 (probability less than 3\%, with random -1/1 outputs). This remains reasonable with respect to usual accuracy of such models.

To evaluate a path on an input $\bm x$, with a ternary result, we need different weights in each comparisons: the values encrypted in the $C_{i,j}^k$ will be 1 or 0 for all the active comparisons except the last node (in red on Figure~\ref{Fig:Tree}), that will contain $\delta$ or 0, where $\delta$ is the length of the paths: if all the comparisons pass, the sum is $2\delta-1$ and corresponds to +1, if all but the last comparison pass, the sum is $\delta-1$ and corresponds to -1, all the other cases will lead to a sum between 0 and $\delta-2$ or $\delta$ and $2\delta-2$ and correspond to 0.
In the following, we will show how the conversion of the sum being $2\delta-1$, $\delta-1$, or anything else can be converted into +1, -1 and 0, respectively, using a verifiable Garbled Circuit.

To make a path with negative polarity, we just invert the comparison in the last node. The inversion by the server will restore the correct value.
We stress that the nodes, after the $C_{i,j}^k$ have been generated, can be randomly permuted to hide which feature is involved in the final node. Furthermore, we remind that since the complexity of our protocol will be independent of the length of the path, any additional comparisons will have no impact to the online phase: we can use them to hide the real structure of the paths and make random guesses of path polarities the best attack for the adversary.

Unfortunately, one cannot rely on path permutations anymore to enforce client privacy once path polarity is used.
Indeed, the server needs to know for which path a score has been computed in order to later involve the correct path polarity $p_i$. 
To overcome this issue, we rely on a secure equality test based on garbled circuits along with an oblivious transfer, to obliviously convert the above sums between 0 and $2\delta-1$ into another ciphertext (under the client key) of +1, -1 or 0.

Hereafter, we use $\bbracket{m}_\sS$ (respectively $\bbracket{m}_\sC$) to denote an encryption of $m$ under the public key of the server (respectively, the client) in order to avoid any confusion. 
The client starts by computing the path score $\boxplus_j \bbracket{S_{i,j}}_\sS$ as previously and masks it with a random value $\alpha_i$ in order to obtain $\bbracket{\beta_i}_\sS$ = $\bbracket{\alpha_i + S_{i}}_\sS$.
As $(S_i=\delta-1) \Leftrightarrow (\alpha_i + \delta-1 =\beta_i)$ and $(S_i=2\delta-1) \Leftrightarrow (\alpha_i + \delta-1 =\beta_i - \delta)$, the client prepares a garbled circuit testing equality of $\alpha_i + \delta-1$ with $\beta_i$ and with $\beta_i - \delta$.
The server computes $\beta_i$ by decrypting $\bbracket{\beta_i}_\sS$ and retrieves the corresponding circuit inputs using an oblivious transfer, thus allowing it to evaluate the aforementioned equality tests, where $\alpha_i + \delta-1$ is the common input from the client, and $\beta_i$ and $\beta_i - \delta$ are the inputs from the server.
Note that in our ElGamal setting, one may use $g^{\beta_i}$ instead of $\beta_i$ for the comparisons, which avoids the server to compute a discrete logarithm during the decryption.

Each of the outcomes of the garbled circuit is mapped to \AHE ciphertexts, under the client key, either $(\bbracket{-1}_\sC,\bbracket{+1}_\sC)$ or $(\bbracket{+1}_\sC,\bbracket{-1}_\sC)$, according to a random choice, along with a pair of \NIZK proving that those ciphertexts are actually encryptions of both $+1$ and $-1$, without revealing the order.

The first equality test labels output +1 in the positive case and -1 otherwise; while the second ones output -1 in the positive case and +1 otherwise: the average of the two values is +1 if the sum is $2\delta-1$; -1 if the sum is $\delta-1$; and 0 otherwise. One can thereafter apply the polarity factor $p_i$ to the ciphertext $\sigma_i$ of the above mean, to restore the real encrypted outcome of the path, under the client key: the server gets, for each path, $\bbracket{+1}_\sC$, $\bbracket{0}_\sC$, or $\bbracket{-1}_\sC$.
All the products $p_i \boxtimes \sigma_i$ are summed up into $\bbracket{S_\Omega}_\sC$, initialized to a random value $\theta$. Hence, $S_\Omega = \theta + \sum_i S_i$, where $S_i\in\{-1,0,+1\}$ is the outcome of each path.
The server will ask the client to help in decrypting this ciphertext, but after having applied a random blinding factor $\zeta\in\bbZ^*_p$, to get back $\zeta(\theta + \sum_i S_i)$. The server can remove $\zeta$ and $\theta$: if the client cheated, the result is random, otherwise this is the number of accepting trees minus the number of rejecting trees. One accepts if this number is between the threshold $\tau$ and the number $T$ of trees. In case the client cheats, the probability to be in this window is less than $T/p$, which is negligible. Again, we stress that discrete logarithms are not needed to check the value is in the window, as the latter is small enough. One can deal with group elements, and not scalars.
The global protocol is described in the appendix.

\subsection{Protocol Security}
\label{S:MaliciousProof}

\paragraph{Correctness and Soundness.} 
The correctness follows the above analysis, where the two equality tests conclude into ciphertexts of -1, +1, or 0, and the path polarity $p_i \in \{-1 , 1 \}$ is thereafter applied to obtain +1 in the accepting case, -1 in the rejecting case, and 0 to ignore the path.

Because of the polarity, we prevent the server from a client arbitrarily choosing the outcome of a path. 
Indeed, in contrast to the honest-but-curious case where expected paths values were zero, the expected value sent to the server (the outcome of the garbled circuit) will depend on the path polarity: +1 if the path has positive polarity, or -1 for the negative polarity, to be an accepting path. 0 values will lead to ignore the path. The paths cannot all be ignored, otherwise, there is no change to be above the threshold $\tau$, hence the two extreme attacks presented before: either the client specifically guesses $\tau$ values to be correct, and set all the other to zero (with a success probability bounded by $2^{-\tau}$), or the client tries non-zero values for all the outputs and the success probability follows a binomial distribution with parameters $(P,1/2)$, where the number of successes must be greater than $(P+\tau)/2$.

The garbled circuits will evaluate the initial scores before polarity, and the unknown polarity bit $p_i$ will restore the exact outcome of the path. A malicious client has no other choice than a random guess of the polarity to fake the output labels of the garbled circuits. He could cheat with a bad encoding of the circuit to bias the output, but only with $+1/-1$ or $-1/+1$ as the output table is proven to contain encryptions of $+1$ and $-1$ with a zero-knowledge proof. But since the player has no idea about the polarity bit $p_i$, the final outcome for the path is $-1$ or $+1$ with identical probability, if positive and negative polarities are balanced.
Hence, alteration of the result of a path, without knowing the polarity, will make the sum closer to $0$. If the threshold is not too close to 0, the probability for the adversary to impersonate the user is negligible. Or at least, the impact of the malicious behaviour of the client on false positive outcome will be small, compared to initial accuracy of the system (in clear).

Of course, another cheating strategy can be sending a false zero-knowledge proof or wrong ciphertexts for the garbled circuit gates. This would lead to failure attacks with a random value in $S_\Omega$ (enforced in the protocol). Similarly, an incorrect decryption of $\bbracket{\widetilde{S_\Omega}}_\sC$ would lead to a random value for $\zeta^{-1} \cdot S_\Omega - \theta$, with the detection probability greater than $1-T/p$, which is overwhelming. This concludes in a reject.

As a consequence, we just have to take care of the accuracy of the model in the clear, for an honest execution, and we will also have to consider the impact of malicious behaviours on the false positive decisions, which is the most critical in the case of continuous authentication.
In some other applications, false negative decisions might be more important to limit (such as for spam detection).


\paragraph{Client Privacy.}
We are still considering the client privacy against an honest-but-curious server. 
However, in this protocol, we no longer use permutations, because of the path polarity.
However, from the client privacy of the oblivious transfers in the garbled circuits, there is no leakage about the $\alpha_i$'s. Then the outcome $S_i$ of the path is encrypted under the client key, which hides it from the server. Eventually, the server only gets the decryption of $\sum_i S_i$, which is the number of accepting trees, the expected result to obtain the classification with confidence score.

\paragraph{Server Privacy.}
Our main goal was the soundness against a malicious client that would try get falsely accepted.
However, this is also important, for our argument of random guess only of the path polarities as the best attack, to show that the adversary cannot learn anything that could help him to make a better guess than at random. Eventually, the client does not get back the decision, so if all the received information looks random, we have proven server privacy.

The first messages received by the client are the encryptions of the comparison gates. Under the indistinguishability of the public-key encryption scheme, they do not leak any information about these gates. Then, the server sends encodings for his inputs $\beta_i$ and $\beta_i-\delta$, which are just keys for the garbled circuit. This does not reveal any information about them to the client.
Eventually, the client receives the encryption of $\widetilde{S_\Omega}$, which is randomized by $\theta$ and $\zeta$. The latter is used to avoid the client to increase his score after decryption while the former completely hides the real value of $\sum_i S_i$

As a consequence, the view of the client does not contain any information about the model, nor the outcome.
Of course, the information to be known to be client is which feature is used in each comparison, in order to use the appropriate $x_{i,j}$. But again, because of possible dummy comparisons, and the random permutations of the comparisons along a path, the client cannot know which gates are real gates, and which gate is the last critical gate.

\section{Performances and Applications}
\label{S:Efficiency}

\subsection{Storage and Bandwidth Costs} \label{S:Performances}
In this section, we describe the storage and bandwidth cost of our protocols.
Let $\lambda^\sR_\OT = |\tilde b|=|\OT.\Encode(b)|$, $ \lambda^\sS_\OT = |\OT.\Compute(\tilde b)|$ and $\lambda_\AHE=|\bbracket{m}_\sS|=|\bbracket{m}_\sC|$.
 
\paragraph{Offline Storage Cost.} 
During the offline phase, the server sends $\bm C$ to the client which requires to store $2^\nu \cdot \delta \cdot P\cdot \lambda_\AHE$ bits.  
Using lifted ElGamal with Elliptic Curves and $p$ over 256 bits as $\kappa=128$, one has $\lambda_\AHE=512$.
With inputs over $\nu \in [2, \ldots, 8]$ bit, depth $\delta\in[2, \ldots, 16]$, and $P$ ranging from 10 to 100, the storage is between a few KB and a few MB.

\paragraph{Honest-but-Curious Bandwidth Cost.}
During the online phase, the client sends $P$ ciphertexts (one for each path score) to the server.
Using the lifted ElGamal \AHE, the message sent by the client is of size $P \cdot 512$ bits (see Table~\ref{T:cost-online}).

\paragraph{Malicious Bandwidth Cost.}
During the online phase, the client sends $P$ tuples, each formed with a ciphertext of size $\lambda_\AHE$ as the path score along with the \AND-gate encodings. 
To reduce communication costs, we use a hash function on the garbled circuit inputs to be compared, with output length $\lambda_\GC$. 
Thus, the client will sent $\lambda_\GC-1$ \AND-gate encodings, the $\lambda_\GC$ input labels (where each label is a hash with size $\kappa_H$) and the transition table consisting in 4-tuple with a garbled circuit output label, a ciphertext, and a ZKP. 

This results for each path in the following bit-length: 
$$
\underbrace{\lambda_\AHE}_{|\bbracket{S_i}|} + \underbrace{\kappa_H \cdot 2(\lambda_\GC-1)}_{|\fC_i|} + \underbrace{\lambda_\GC \cdot \kappa_H}_{|\fI_{\alpha_i}^\sG|} + \underbrace{2 \cdot (2\kappa_H + \lambda_\AHE)}_{|\bm \fT_i|})
$$

which is $3(\lambda_\AHE + \lambda_\GC \cdot \kappa_H)+2 \kappa_H $.
With $\kappa_H=256,\lambda_\GC=64$ and $\lambda_\AHE=512$, the first client's message can be expressed as $51200P$ bits, i.e. 6.25 KB per path.
During the oblivious transfers, the server encodes $P \times |\beta_i|=P \cdot \lambda_\GC$ bits. 
As a consequence, he sends to the client a message of $P \cdot \lambda_\GC \cdot \lambda^\sR_\OT$ bits. 
Using classical ElGamal as \PKE in the oblivious transfer, one has $\lambda^\sR_\OT=512$ and $\lambda^\sS_\OT=1024$ resulting in a 4 KB long message for each path.
The client responds with a $P \cdot \lambda_\GC \cdot \lambda^\sS_\OT$ bit long message corresponding to 8 KB per path.
Then, the server sends the encrypted randomized score to the client who returns the plaintext value he retrieves when decrypting.
Table \ref{T:cost-online} shows the total bandwidth cost for each party in both protocols, according to the number $P$ of paths.

\begin{table}[h] \small
    \caption{Communications During the Online Phase.  \label{T:cost-online}}
\centering        \begin{tabular}{|c|c|c|c|c|} \hline
            $P$  & 50 & 100 & 150 & 200\\ \hline
            \multicolumn{5}{|c|}{\cellcolor{gray!20} Honest-but-Curious} \\ \hline
            Client & 3.13 KB & 6.25 KB & 9.38 KB & 12.5 KB \\ \hline
            Server & \multicolumn{4}{c|}{0 bit} \\ \hline
            \multicolumn{5}{|c|}{ \cellcolor{gray!20} Malicious} \\ \hline
            Client & 712.5 KB & 1.4 MB & 2.1MB & 2.8MB\\ \hline
            Server & 200 KB & 400 KB & 600KB & 800KB\\ \hline
        \end{tabular} \vspace*{-1em}
\end{table}

\subsection{Application to Continuous Authentication and Spam Filtering}
\label{S:Applications}
We run our tests in Python with the \textsf{scikit-learn} library, using 75\% of the dataset as the training set and the remaining 25\% as the testing set. We optimize the training with the Orthogonal Matching Pursuit Algorithm \cite{OMP}: we generate 100 times more trees than expected. We then apply the OMP algorithm on the global set, such that the outcome is the best linear combination with the expected number of trees.

  \begin{table}[b] \small
    \caption{Accuracy on our continuous authentication Database \label{T:CA-M}}
      \centering
      \begin{tabular}{|c|c||c|c|c||c|c|c|}
      \cline{1-8}
      \multicolumn{2}{|c||}{$\Gamma$} & \multicolumn{3}{c||}{50\%}  & \multicolumn{3}{c||}{55\%} \\ \hline \hline
      $T$                 & $\delta$  & FPR   & FNR   & F1 Score    & FPR   & FNR   & F1 Score   \\ \hline
      \multirow{2}{*}{10} & 6         & 0.02  & 0.15  & 0.92        & 0.02  & 0.16  & 0.91       \\ \cline{2-8} 
                          & 8         & 0.01  & 0.19  & 0.91        & 0.01  & 0.19  & 0.91       \\ \hline
      \multirow{2}{*}{25} & 6         & 0.04  & 0.13  & 0.92        & 0.03  & 0.15  & 0.92       \\ \cline{2-8} 
                          & 8         & 0.02  & 0.14  & 0.92        & 0.02  & 0.17  & 0.91       \\ \hline
      \end{tabular}
      \begin{tabular}{|c|c|c||c|c|c|}
        \cline{1-6}
       \multicolumn{3}{|c||}{60\%} & \multicolumn{3}{c|}{65\%} \\ \hline \hline
      FPR   & FNR   & F1 Score    & FPR   & FNR   & F1 Score  \\ \hline
      0.01  & 0.23  & 0.89        & 0.01  & 0.23  & 0.89      \\ \hline
      0.01  & 0.25  & 0.88        & 0.01  & 0.26  & 0.88      \\ \hline
      0.02  & 0.2   & 0.9         & 0.01  & 0.23  & 0.89      \\ \hline
      0.01  & 0.24  & 0.89        & 0.01  & 0.26  & 0.88      \\ \hline
      \end{tabular} 
    \end{table}

\begin{table}\small
  \caption{Accuracy on the spambase Database \label{T:spambase-M}}
   \centering
    \begin{tabular}{|c|c||c|c|c||c|c|c|}
      \cline{1-8}
      \multicolumn{2}{|c||}{$\Gamma$} & \multicolumn{3}{c||}{50\%}  & \multicolumn{3}{c||}{55\%} \\ \hline \hline
      $T$                 & $\delta$  & FPR   & FNR   & F1 Score    & FPR   & FNR   & F1 Score  \\  \hline
        \multirow{2}{*}{10}      & 2  & 0.02  & 0.24  & 0.88        & 0.01  & 0.28  & 0.87    \\    \cline{2-8}
                                & 4  & 0.03  & 0.20  & 0.89        & 0.04  & 0.18  & 0.90     \\   \cline{1-8}
        \multirow{2}{*}{25}      & 2  & 0.03  & 0.24  & 0.88        & 0.02  & 0.25  & 0.88    \\    \cline{2-8}
                                & 4  & 0.04  & 0.19  & 0.89        & 0.04  & 0.21  & 0.88     \\   \hline
       \end{tabular}
         \begin{tabular}{|c|c|c||c|c|c|}
             \cline{1-6}
            \multicolumn{3}{|c||}{60\%} & \multicolumn{3}{c|}{65\%} \\ \hline \hline
            FPR   & FNR   & F1 Score    & FPR   & FNR   & F1 Score  \\ \hline
            0.02  & 0.24  & 0.88        &   0.01  & 0.45  & 0.81    \\ \hline
            0.02  & 0.27  & 0.87        &   0.01  & 0.30  & 0.86    \\ \hline
            0.01  & 0.47  & 0.80        &   0.00  & 0.56  & 0.78    \\ \hline
            0.02  & 0.23  & 0.89        &   0.01  & 0.34  & 0.85    \\ \hline
          \end{tabular}
  \end{table}

We first deal with continuous authentication. We used an internal database of 20531 samples splitted in 35 profiles built with 222 features. In this context, low \emph{False Positive Rate} (FPR) is privileged to low \emph{False Negative Rate} (FNR), since it is preferable to ask the client to use a second authentication factor rather than being impersonated. Moreover, high accuracy for each test is not required, as multiple tests will amplify the quality. Table~\ref{T:CA-M} shows the mean results ont the 35 profiles (where, for a given profile, all other profiles are considered as imposter), depending on the number of paths $P$ and the depth of the model $\delta$, while $\nu$ is set to 6, leading to 64 ciphertexts for each comparison, stored by the client. Also, we consider several values for the acceptation threshold ($\Gamma$) (which equals 50\% by default, for the simple majority). 

We compute the FPR and FNR, then the accuracy is defined as the F1-score (defined by $(1-\mathsf{FPR})/(1+(\mathsf{FNR}-\mathsf{FPR})/2)$). Random decision would lead to an accuracy of 1/2, and perfect filter should have F1-score equal to 1. 
We determine the best accuracy depending on those parameters in Table~\ref{T:CA-M}. For the honest-but-curious security setting, there is no constraint on the threshold, while against malicious clients, the higher the threshold is, the higher the security level is against active impersonation attempts.

Secondly, we worked on the spambase database \cite{UCI:spambase} (with 4601 samples $\times$ 57 features) which determines if an email should be considered as spam or not. Results are shown in Table~\ref{T:spambase-M}

\section{Conclusion}
\label{S:Conclusion}

In this paper, we proposed new constructions to securely evaluate decision forests with two output classes.
As we targeted applications where the interactions between the client and the server should be as low as possible (both in term of number of rounds and online bandwidth cost), we designed two-steps protocols in which some part of the computation can be performed offline.
This introduces an interesting trade-off between the storage and the number of rounds during the online step of the protocol.



\bibliographystyle{alpha}
\bibliography{cryptobib/abbrev3.bib,cryptobib/crypto.bib,other.bib}

\newcommand{\etalchar}[1]{$^{#1}$}
\begin{thebibliography}{AMPR14}

\bibitem[AMPR14]{EC:AMPR14}
Arash Afshar, Payman Mohassel, Benny Pinkas, and Ben Riva.
\newblock Non-interactive secure computation based on cut-and-choose.
\newblock In Phong~Q. Nguyen and Elisabeth Oswald, editors, {\em
  EUROCRYPT~2014}, volume 8441 of {\em {LNCS}}, pages 387--404. Springer,
  Heidelberg, May 2014.

\bibitem[BFK{\etalchar{+}}09]{ESORICS:BFKLSS09}
Mauro Barni, Pierluigi Failla, Vladimir Kolesnikov, Riccardo Lazzeretti,
  Ahmad-Reza Sadeghi, and Thomas Schneider.
\newblock Secure evaluation of private linear branching programs with medical
  applications.
\newblock In Michael Backes and Peng Ning, editors, {\em ESORICS~2009}, volume
  5789 of {\em {LNCS}}, pages 424--439. Springer, Heidelberg, September 2009.

\bibitem[BMR90]{STOC:BeaMicRog90}
Donald Beaver, Silvio Micali, and Phillip Rogaway.
\newblock The round complexity of secure protocols (extended abstract).
\newblock In {\em 22nd ACM STOC}, pages 503--513. {ACM} Press, May 1990.

\bibitem[BPSW07]{CCS:BPSW07}
Justin Brickell, Donald~E. Porter, Vitaly Shmatikov, and Emmett Witchel.
\newblock Privacy-preserving remote diagnostics.
\newblock In Peng Ning, Sabrina {De Capitani di Vimercati}, and Paul~F.
  Syverson, editors, {\em ACM CCS 2007}, pages 498--507. {ACM} Press, October
  2007.

\bibitem[BPTG15]{NDSS:BPTG15}
Raphael Bost, Raluca~Ada Popa, Stephen Tu, and Shafi Goldwasser.
\newblock Machine learning classification over encrypted data.
\newblock In {\em NDSS~2015}. The Internet Society, February 2015.

\bibitem[DDH{\etalchar{+}}16]{EPRINT:DDHKNN16}
Martine {De Cock}, Rafael Dowsley, Caleb Horst, Raj Katti, Anderson C.~A.
  Nascimento, Stacey~C. Newman, and Wing-Sea Poon.
\newblock Efficient and private scoring of decision trees, support vector
  machines and logistic regression models based on pre-computation.
\newblock Cryptology ePrint Archive, Report 2016/736, 2016.
\newblock \url{https://eprint.iacr.org/2016/736}.

\bibitem[DPB18]{PROVSEC:DupPoiBid18}
Aur{\'e}lien Dupin, David Pointcheval, and Christophe Bidan.
\newblock On the leakage of corrupted garbled circuits.
\newblock In Joonsang Baek, Willy Susilo, and Jongkil Kim, editors, {\em
  ProvSec 2018}, volume 11192 of {\em {LNCS}}, pages 3--21. Springer,
  Heidelberg, October 2018.

\bibitem[EGL82]{C:EveGolLem82}
Shimon Even, Oded Goldreich, and Abraham Lempel.
\newblock A randomized protocol for signing contracts.
\newblock In David Chaum, Ronald~L. Rivest, and Alan~T. Sherman, editors, {\em
  CRYPTO'82}, pages 205--210. Plenum Press, New York, USA, 1982.

\bibitem[{ElG}84]{C:ElGamal84}
Taher {ElGamal}.
\newblock A public key cryptosystem and a signature scheme based on discrete
  logarithms.
\newblock In G.~R. Blakley and David Chaum, editors, {\em CRYPTO'84}, volume
  196 of {\em {LNCS}}, pages 10--18. Springer, Heidelberg, August 1984.

\bibitem[HRFS99]{UCI:spambase}
Mark Hopkins, Erik Reeber, George Forman, and Jaap Suermondt.
\newblock {\em Spambase Data Set}, 1999.
\newblock \url{http://archive.ics.uci.edu/ml/datasets/Spambase/}.

\bibitem[KS08]{ICALP:KolSch08}
Vladimir Kolesnikov and Thomas Schneider.
\newblock Improved garbled circuit: Free {XOR} gates and applications.
\newblock In Luca Aceto, Ivan Damg{\aa}rd, Leslie~Ann Goldberg, Magn{\'u}s~M.
  Halld{\'o}rsson, Anna Ing{\'o}lfsd{\'o}ttir, and Igor Walukiewicz, editors,
  {\em ICALP 2008, Part~II}, volume 5126 of {\em {LNCS}}, pages 486--498.
  Springer, Heidelberg, July 2008.

\bibitem[Lin13]{C:Lindell13}
Yehuda Lindell.
\newblock Fast cut-and-choose based protocols for malicious and covert
  adversaries.
\newblock In Ran Canetti and Juan~A. Garay, editors, {\em CRYPTO~2013,
  Part~II}, volume 8043 of {\em {LNCS}}, pages 1--17. Springer, Heidelberg,
  August 2013.

\bibitem[LP07]{EC:LinPin07}
Yehuda Lindell and Benny Pinkas.
\newblock An efficient protocol for secure two-party computation in the
  presence of malicious adversaries.
\newblock In Moni Naor, editor, {\em EUROCRYPT~2007}, volume 4515 of {\em
  {LNCS}}, pages 52--78. Springer, Heidelberg, May 2007.

\bibitem[LR14]{C:LinRiv14}
Yehuda Lindell and Ben Riva.
\newblock Cut-and-choose {Yao}-based secure computation in the online/offline
  and batch settings.
\newblock In Juan~A. Garay and Rosario Gennaro, editors, {\em CRYPTO~2014,
  Part~II}, volume 8617 of {\em {LNCS}}, pages 476--494. Springer, Heidelberg,
  August 2014.

\bibitem[MF06]{PKC:MohFra06b}
Payman Mohassel and Matthew Franklin.
\newblock Efficiency tradeoffs for malicious two-party computation.
\newblock In Moti Yung, Yevgeniy Dodis, Aggelos Kiayias, and Tal Malkin,
  editors, {\em PKC~2006}, volume 3958 of {\em {LNCS}}, pages 458--473.
  Springer, Heidelberg, April 2006.

\bibitem[MZ93]{OMP}
S.G. Mallat and Zhifeng Zhang.
\newblock Matching pursuits with time-frequency dictionaries.
\newblock {\em IEEE Transactions on Signal Processing}, 41(12):3397--3415,
  1993.

\bibitem[TKK19]{PoPETS:TueKerKat19}
Anselme Tueno, Florian Kerschbaum, and Stefan Katzenbeisser.
\newblock Private evaluation of decision trees using sublinear cost.
\newblock {\em {PoPETs}}, 2019(1):266--286, January 2019.

\bibitem[TMZC17]{ESORICS:TMZC17}
Raymond K.~H. Tai, Jack P.~K. Ma, Yongjun Zhao, and Sherman S.~M. Chow.
\newblock Privacy-preserving decision trees evaluation via linear functions.
\newblock In Simon~N. Foley, Dieter Gollmann, and Einar Snekkenes, editors,
  {\em ESORICS~2017, Part~II}, volume 10493 of {\em {LNCS}}, pages 494--512.
  Springer, Heidelberg, September 2017.

\bibitem[WFNL16]{PoPETS:WFNL16}
David~J. Wu, Tony Feng, Michael Naehrig, and Kristin~E. Lauter.
\newblock Privately evaluating decision trees and random forests.
\newblock {\em {PoPETs}}, 2016(4):335--355, October 2016.

\bibitem[WMK17]{EC:WanMalKat17}
Xiao Wang, Alex~J. Malozemoff, and Jonathan Katz.
\newblock Faster secure two-party computation in the single-execution setting.
\newblock In Jean-S{\'{e}}bastien Coron and Jesper~Buus Nielsen, editors, {\em
  EUROCRYPT~2017, Part~III}, volume 10212 of {\em {LNCS}}, pages 399--424.
  Springer, Heidelberg, April~/~May 2017.

\bibitem[Yao86]{FOCS:Yao86}
Andrew Chi-Chih Yao.
\newblock How to generate and exchange secrets (extended abstract).
\newblock In {\em 27th FOCS}, pages 162--167. {IEEE} Computer Society Press,
  October 1986.

\bibitem[ZRE15]{EC:ZahRosEva15}
Samee Zahur, Mike Rosulek, and David Evans.
\newblock Two halves make a whole - reducing data transfer in garbled circuits
  using half gates.
\newblock In Elisabeth Oswald and Marc Fischlin, editors, {\em EUROCRYPT~2015,
  Part~II}, volume 9057 of {\em {LNCS}}, pages 220--250. Springer, Heidelberg,
  April 2015.

\end{thebibliography}

\newpage
\appendix
\section{Auxiliary Material}

We first give a few more formal description of some advanced tools, to be used the global protocol, and then more details about the global security decision forest evaluation for malicious clients.

\subsection{Advanced Primitives}
\paragraph{Garbled Circuits}
From our modeling of them, we can define Garbled Circuits with two algorithms $(\Generate, \Eval)$, in addition to an \OT, as follows:
\begin{itemize}
 \item $\Generate(\mathcal C, \alpha)$: Given a circuit $\cC$ and the generator input $\alpha$, returns:
   \subitem The ciphertexts corresponding to the gates of $\cC$: $\fC$
   \subitem The input labels corresponding to the generator input $\alpha$: $\fI^\sG_\alpha$
   \subitem The input labels $\fI^\sE$ corresponding to the possible inputs of the evaluator (but not published)
   \subitem The transition table mapping the output labels with arbitrary values: $\fT$;
 \item \OT~Execution: on input $\fI^\sE$ from the generator (as sender) and the input $\beta$ of the evaluator (as receiver), for the latter to receive the input labels $\fI^\sE_\beta$;
 \item $\Eval(\fC, \fI^\sG_\alpha,\fI^\sE_\beta,\fT)$: From the garbled circuit $\fC$, input labels $\{\fI^\sG_\alpha, \fI^\sE_\beta\}$ and transition table $\fT$, retrieve $\fO_b \in (\fO_0, \fO_1)$ by evaluating the garbled circuit $\fC$ with input labels $\fI^\sG_\alpha, \fI^\sE_\beta$ and return the value mapped by $\fO_b$ in $\fT$. If the garbled circuit evaluation fails (the output value is not in $(\fO_0, \fO_1)$, return $\perp$, as a failure outcome.
\end{itemize}
The generator should not learn any information about the evaluator input $\alpha$, while the evaluator should not learn any information about the generator input $\beta$, except what the result reveals.
These privacy notions rely on the privacy of the \OT: with receiver-privacy against a malicious sender, we then get the privacy of the evaluator against a malicious generator, and with sender-privacy against an honest-but-curious receiver, we get the privacy of the generator against an honest-but-curious evaluator.
Another important notion is of course the correct evaluation of the function, a.k.a. the soundness. Since the evaluator is interested in the correct result, the attack can come from the generator that provides a wrong encoding of the circuit. A classical technique to avoid incorrect circuits is based on a costly cut-and-choose. One of our contributions is an efficient alternative in our particular case.

\paragraph{Zero-Knowledge Proofs} 
A \emph{Zero-Knowledge Proof} (\ZKP) is an interactive protocol between a prover $P$, who wants to prove to a verifier $V$, that a given statement belongs to a language, without leaking any information about the witness. It can be made non-interactiveWe will consider Non-Interactive Zero-Knowledge proofs (\NIZK):
\begin{itemize}
  \item $\zkgen(x,L, w)$ outputs $\pi_x$, a zero-knowledge proof of the statement $(x \in L)$, using the witness $w$;
  \item $\zkverif(x,L,\pi_x)$ verifies if $\pi_x$ is a correct proof of the statement $(x \in L)$. Return \accept if true, \reject otherwise;
  \item $\zksim(x,L)$ simulates a proof $\pi_x$ of any (possibly false) statement $(x\in L)$ without any witness.
\end{itemize} 
A \NIZK requires three properties:
\begin{itemize}
  \item \textbf{Completeness}: $\zkgen$ always generate an acceptable proof when $x \in L$;
  \item \textbf{Soundness}: no adversary can generate an acceptable proof when $x \notin L$, but with negligible probability;
  \item \textbf{Zero-Knowledge}: using possibly a different setup, $\zksim(x,L)$ generates proofs that are indistinguishable to proofs generated by $\zkgen(x,L,w)$, on valid statements but without the witness.
\end{itemize}

\subsection{Detailed Protocol} 
In this section we give a detailed description of the protocol in the malicious client setting. We first describe the steps executed by the server then the steps done by the client. We recall that $\bbracket{m}_\sS=\AHE.\enc(\pk_\sS,m)$ and $\bbracket{m}_\sC=\AHE.\enc(\pk_\sC,m)$. 

\paragraph{a) Client steps} ~\\
input : $P, \delta, \nu, \bm x = (x_{i,j})_{i \in [P], j\in[\delta]}$ \\
roles: \GC: Generator , \OT: Sender\\[1em]
\begin{tabular}{|p{\linewidth}|} \hline
  $(\pk_\sC,\sk_\sC) = \AHE.\keygen(1^\kappa)$ \\
  \multicolumn{1}{|c|}{\cellcolor{gray!20} $(\pk_\sC, \sk_\sC) = \Keygen_\sC(1^\kappa)$} \\ \hline
\end{tabular} \\
\begin{tabular}{|p{\linewidth}|} \hline		
  for $i \in [1 \ .. \ P]$: $\alpha_i \sampled \mathbb Z_p$ \\
  \quad $(\fC_i, \fI^\sG_{\alpha_i}, \fI_i^\sE, \fT_i) \gets \GC.\Generate(\pk_\sC, \mathcal C_{EQ}, \alpha_i )$ \\ 
  \multicolumn{1}{|c|}{\cellcolor{gray!20} $(\bm \alpha, \fC, \fI^\sG_{\bm \alpha}, \fI^\sE, \fT) = \Initialization(\pk_\sC)$} \\  \hline
\end{tabular} \\
\begin{tabular}{|p{\linewidth}|} \hline
  for $i \in [1 \ .. \ P]$: \\
  $\quad$ for  $j \in [1 \ .. \ \delta]:$ $\bbracket{S_{i,j}}_\sS= C_{i,j}^{x_{i,j}} $ \\
      $\quad \bbracket{\beta_i}_\sS=\bbracket{\alpha_i}_\sS\scalebox{1.2}{$\boxplus$} \left(\scalebox{1.5}{$\boxplus$}_j\bbracket{S_{i,j}}_\sS \right) $ \\ 
      \multicolumn{1}{|c|}{\cellcolor{gray!20}$\bm{\bbracket{\beta}}_\sS = \EvalPaths(\pk_\sS, \bm x, \bm \alpha, \bm C)$} \\ \hline
  \end{tabular} \\ 
\begin{tabular}{|p{\linewidth}|} \hline 
  for $i \in [1 \ .. \ P]$, for $k \in [1 \ .. \ \lambda_\AHE]:$ \\
  \qquad $\widetilde{\fI^\sE_{i,k}} \gets \OT.\Compute(\fI_i^\sE, \widetilde \beta_i^k) $ \\ 
      \multicolumn{1}{|c|}{\cellcolor{gray!20}$\widetilde{\fI^\sE} = \ComputeOT(\fI^\sE, \bm{\widetilde \beta})$} \\ \hline
\end{tabular} \\
\begin{tabular}{|p{\linewidth}|} \hline
  $\widetilde{S_\Omega} \gets \AHE. \dec(\sk_\sC,\bbracket{\widetilde{S_\Omega}}_\sC)$ \\  
  \multicolumn{1}{|c|}{\cellcolor{gray!20} $\widetilde{S_\Omega}=\DecryptScore(\sk_\sC, \bbracket{\widetilde{S_\Omega}}_\sC)$} \\ \hline 
\end{tabular} \\

\paragraph{b) Server steps} ~\\
input: $\mathcal{M} = (P, \delta, \nu, \tau, (p_i, t_{i, j}, v_{i, j})_{i \in [P], j \in [\delta]})$ \\
roles: \GC: Evaluator , \OT: Receiver \\ [1em]
\begin{tabular}{|p{\linewidth}|} \hline
  $(\pk_\sS,\sk_\sS) = \AHE.\keygen(1^\kappa)$ \\
  \multicolumn{1}{|c|}{\cellcolor{gray!20} $(\pk_\sS, \sk_\sS) = \Keygen_\sS(1^\kappa)$} \\ \hline
\end{tabular} \\
\begin{tabular}{|p{\linewidth}|} \hline
  for $i \in [1 \ .. \ P]$: $\pi \gets \mathfrak{S}_\delta$ \\
    \quad for $k \in [0\ .. \ 2^\nu-1]$, for $j \in [1 \ .. \ \delta-1]$: \\
    \quad \begin{tabular}{r@{ }c@{ }l}
        $C_{i,\pi(j)}^k$ & = & $\left\{
        \begin{array}{l}
          \bbracket{1-v_{i,j}}_\sS \mbox{ if } k \leq t_{i,j} \\
          \bbracket{v_{i,j}}_\sS \mbox{ otherwise.}
        \end{array}
        \right.$ \\
        $C_{i,\pi(\delta)}^k$ & = & $\left\{
      \begin{array}{l}
        \bbracket{((1+p_i)/2 - v_{i,\delta} p_i) \cdot \delta}_\sS \mbox{ if } k \leq t_{i,j} \\
        \bbracket{((1-p_i)/2 + v_{i,\delta} p_i) \cdot \delta}_\sS \mbox{ otherwise.}
      \end{array}
    \right.$
    \end{tabular} \\
    \multicolumn{1}{|c|}{\cellcolor{gray!20} $\bm C = \EncodeModel(\pk_\sS, \mathcal{M})$} \\ \hline
\end{tabular} \\  
\begin{tabular}{|p{\linewidth}|} \hline 
  for $i \in [1 \ .. \ P]:$ $\beta_i=\AHE.\dec(\sk_\sS,\bbracket{\beta_i}_\sS)$ \\ 
  \quad for $k \in [1 \ .. \ \lambda_\AHE]:$ $\widetilde \beta_{i}^k=\OT.\Encode(\beta_i[k])$ \\ 
  \multicolumn{1}{|c|}{\cellcolor{gray!20}$\widetilde{\bm \beta} = \EncodeOT(\sk_\sS, \bm{ \bbracket \beta}_\sS)$} \\   \hline
\end{tabular} \\  
\begin{tabular}{|p{\linewidth}|} \hline
  $\theta \sampled \bbZ_p$;
  $\bbracket{S_\Omega}_\sC \gets \bbracket{\theta}_\sC$ \\
  for $i \in [1 \ .. \ P]:$ \\
    \quad for $k \in [1 \ .. \ \lambda_\AHE]: \fI_{\beta_i^k}^\sE \gets \OT.\Decode(\widetilde{\fI^\sE_{i,k}})$ \\
    \quad if $\GC.\Eval(\fC_i, \fI_{\alpha_i}^\sG, \fI_{\beta_i}^\sE, \fT_i)=\perp$: \\[-.4em]
      \qquad $\theta' \sampled \bbZ_p$; $\bbracket{S_\Omega}_\sC \gets \bbracket{\theta'}_\sC$ \\
    \quad else:  $((\sigma_0 , \pi_0),(\sigma_1, \pi_1)) \gets \GC.\Eval(\fC_i, \fI_{\alpha_i}^\sG, \fI_{\beta_i}^\sE, \fT_i)$ \\
    \quad if $\zkverif(((\sigma_0 , \pi_0),(\sigma_1, \pi_1)))=\accept$: \\
      \qquad $\bbracket{S_\Omega}_\sC \gets \bbracket{S_\Omega}_\sC \boxplus (p_i \boxtimes (\sigma_0 \boxplus \sigma_1)) $ \\ 
    \quad else: $\theta' \sampled \bbZ_p$; $\bbracket{S_\Omega}_\sC \gets \bbracket{\theta'}_\sC$ \\
  $\zeta \sampled \bbZ^*_p$;
  $\bbracket{\widetilde{S_\Omega}}_\sC \gets \zeta \boxtimes \bbracket{S_\Omega}_\sC$ \\
  \multicolumn{1}{|c|}{\cellcolor{gray!20} $(\bbracket{\widetilde{S_\Omega}}_\sC , \theta, \zeta) = \ComputeScore(\pk_\sC, \mathcal{M}, \fC, \fI_{\bm \alpha}^\sG, \widetilde{\fI^\sE}, \fT)$} \\ \hline 
\end{tabular} \\
\begin{tabular}{|p{\linewidth}|} \hline
  $S \gets \zeta^{-1} \cdot \widetilde{S_\Omega} - \theta$ \\
    if $S \in [\tau, T]: \result = 1$;
    else: $\result = 0$ \\
    \multicolumn{1}{|c|}{\cellcolor{gray!20}$\result = \ComputeResult(\mathcal{M}, \theta, \zeta, \widetilde{S_\Omega})$} \\ \hline
  \end{tabular} 

\paragraph{c) Protocol} We now describe the protocol execution: \\
\begin{tabular}{lcr} 
  \rowcolor{gray!30}\multicolumn{3}{c}{Offline phase} \\
  && $\Keygen_\sS()$ \\
  &$\sendleft{\pk_\sS,~\bm C}{22mm}$& $\EncodeModel()$ \\ 
  \rowcolor{gray!30}\multicolumn{3}{c}{Online phase} \\
  $\Keygen_\sC()$ && \\
  $\Initialization()$ \\ 
  $\EvalPaths()$ & $\sendright{\pk_\sC,\ \bm \beta,\ (\fC,\fI_\alpha^\sG, \fT)}{22mm}$ & \\
  &$\sendleft{\widetilde{\bm \beta}}{22mm}$& $\Encode\OT()$ \\
  $\Compute\OT()$ & $\sendright{\widetilde{\fI^\sE}}{22mm}$&\\
  &$\sendleft{\bbracket{\widetilde{S_\Omega}}}{22mm}$& $\ComputeScore()$ \\
  $ \DecryptScore()$ & $\sendright{\widetilde{S_\Omega}}{22mm} $\\
  &&$\ComputeResult()$   
\end{tabular}

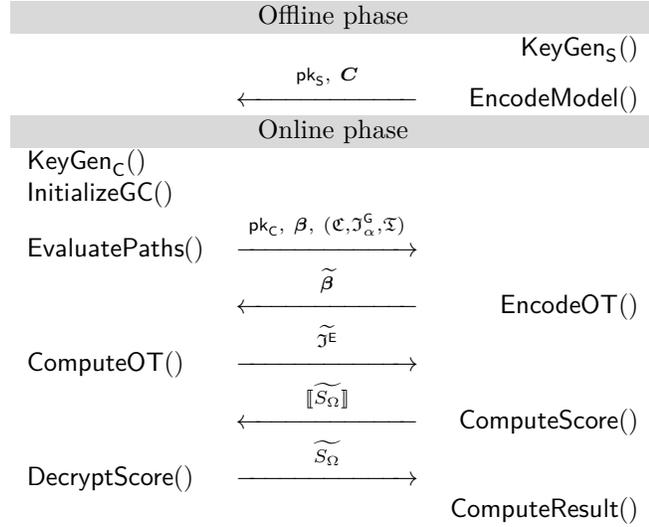
\end{document}